\documentclass[fleqn,12pt]{wlscirep}
\usepackage{wrapfig, framed, caption}
\usepackage[utf8]{inputenc}
\usepackage{heuristica}
\usepackage{graphicx}
\usepackage{wrapfig, framed, caption}
\usepackage{tcolorbox}
\usepackage{dirtytalk}
\usepackage{xr}

\makeatletter
\newcommand*{\addFileDependency}[1]{
  \typeout{(#1)}
  \@addtofilelist{#1}
  \IfFileExists{#1}{}{\typeout{No file #1.}}
}
\makeatother
 
\newcommand*{\myexternaldocument}[1]{%
    \externaldocument[ext1-]{#1}%
    \addFileDependency{#1.tex}%
    \addFileDependency{#1.aux}%
}
\myexternaldocument{SI}

\usepackage{setspace}
\usepackage[normalem]{ulem}
\useunder{\uline}{\ulined}{}
\usepackage{xparse}
\newsavebox{\fminipagebox}
\NewDocumentEnvironment{fminipage}{m O{\fboxsep}}
 {\par\kern#2\noindent\begin{lrbox}{\fminipagebox}
  \begin{minipage}{#1}\ignorespaces
 \end{minipage}\end{lrbox}%
  \makebox[#1]{%
    \kern\dimexpr-\fboxsep-\fboxrule\relax
    \fbox{\usebox{\fminipagebox}}%
    \kern\dimexpr-\fboxsep-\fboxrule\relax
  }\par\kern#2
 }

\title{Elites, communities and the limited benefits of mentorship in electronic music}

\author[1]{Mil\'an Janosov}
\author[1]{Federico Musciotto}
\author[1]{Federico Battiston}
\author[1, 2, 3,*]{Gerardo Iñiguez}

\affil[1]{Department of Network and Data Science, Central European University, Budapest, 1051, Hungary}
\affil[2]{Department of Computer Science, Aalto University School of Science, Aalto, 00076, Finland}
\affil[3]{IIMAS, Universidad Nacional Auton\'{o}ma de M\'{e}xico, Ciudad de M\'{e}xico, 01000, Mexico}

\affil[*]{iniguezg@ceu.edu}

\begin{document}

\maketitle


\section*{Abstract}
{\small
While the emergence of success in creative professions, such as music, has been studied extensively, the link between individual success and collaboration is not yet fully uncovered. Here we aim to fill this gap by analyzing longitudinal data on the co-releasing and mentoring patterns of popular electronic music artists appearing in the annual Top 100 ranking of DJ Magazine. We find that while this ranking list of popularity publishes 100 names, only the top 20 is stable over time, showcasing a lock-in effect on the electronic music elite. Based on the temporal co-release network of top musicians, we extract a diverse community structure characterizing the electronic music industry. These groups of artists are temporally segregated, sequentially formed around leading musicians, and represent changes in musical genres. We show that a major driving force behind the formation of music communities is mentorship: around half of musicians entering the top 100 have been mentored by current leading figures before they entered the list. We also find that mentees are unlikely to break into the top 20, yet have much higher expected best ranks than those who were not mentored. This implies that mentorship helps rising talents, but becoming an all-time star requires more. Our results provide insights into the intertwined roles of success and collaboration in electronic music, highlighting the mechanisms shaping the formation and landscape of artistic elites in electronic music.
}

\vspace{0.5cm}
{\small {\bf Keywords}: success, music, rank dynamics, communities, mentorship}

\section{Introduction}

Throughout history, music has been one of the most powerful forms of culture and identity expression. Music is typically not the product of an individual mind, but the result of a collaborative effort involving people with diverse backgrounds and behaviors. The world of musicians is, therefore, a complex social ecosystem, showcasing a myriad of genres, trends, tools, and audiences. In the era of big data~\cite{lazer2009life, clauset2017data, fortunato2018science, barabasi2018formula, li2019nobel}, many researchers have turned their attention to creative fields from science to music and attempted to quantify exceptional success. Some of these works, originated in developmental psychology, focus on uncovering the roots of individual success and career trajectories in music~\cite{simonton1989swan, simonton1991emergence,ericsson2006influence, wang2019gender}, like the role of forbidden triads or the relational field in jazz~\cite{vedres2017forbidden, kirschbaum2017trajectory}, while others aim to give general, data-driven explanations on how individual careers evolve~\cite{simonton1984creative, sinatra2016quantifying, liu2018hot}. Some researchers have attempted to capture large-scale features of the musical world, such as extracting collaboration and community structure, or identifying genres of various scenes like classical music, jazz, and the Rolling Stone Magazine’s list of ‘500 Greatest Albums of All Time’~\cite{gleiser2003community,burke2014clam, budner2016collaboration,park2015topology}. More recent works have also analyzed the changes of trends and fashion cycles in music over time~\cite{klimek2019fashion, youngblood2019cultural,park2019global}.

Still, a clear connection between the success of individuals and their role in the music scenes' social fabric is lacking. Here we aim to fill this gap by investigating the well-defined ecosystem of artists working on electronic music. During the past two decades, electronic music transitioned from the outskirts of music to become one of its most popular fields. Yet surprisingly, electronic music has only produced a handful of stars typically performing in front of tens of thousands of people, while the majority of disc jockeys (DJs) and producers remain unknown. Hence, our goal is to better understand how superstar DJs and producers (since a large fraction of DJs also act as producers) emerge by analyzing the interplay between individual success, quantified in terms of the top 100 DJs' ranking list from 1997 to 2018  (curated by DJ Magazine~\cite{djmag, amf}), and the underlying collaborations captured based on Discogs~\cite{discogs,hartnett2015discogs}. We also detect the structure and the dynamics of the various communities in the artists' co-release network and trace the effects of mentorship on young musicians. In fact, mentorship is usually the door through which new talents enter social environments whose activity is based on skills that require long traineeships, and its effect has already been tracked within academia\cite{sekara2018chaperone}. 

We first analyze the dynamics of the DJ ranking list~\cite{blumm2012dynamics, cocho2015rank, morales2016generic, sanchez2018trajectory, morales2018rank} in order to capture the most stable subset of star DJs and thus define the superstars appearing during 22 years of available data. We then connect this dynamics with the collaboration network among musicians by looking at the structure of communities and characterizing them with respect to their prevailing subgenre (such as house and techno) and leading figures. Finally, we provide a definition for mentorship and study its consequences on the careers of young DJs by looking at the success trajectories of mentees in the collaboration network.


\section{Data}
\label{Sec:data}

We collect the annual top 100 ranking list of DJs from the official website of DJ Magazine~\cite{djmag} and related sources~\cite{djranking2, djranking3}. This ranking list, officially announced at the Amsterdam Music Festival in recent years~\cite{amf}, is based on a yearly poll filled out by several million people~\cite{top100}, and is traditionally considered to be a good proxy for DJ popularity. During 1997-2018, 540 DJs have managed to enter this elite club of electronic music artists. We complement our ranking dataset with information extracted from Discogs~\cite{discogs, hartnett2015discogs}, an online crowd-sourced music discography platform that lists the production of 46,063 artists active on electronic music, comprising 1,103,769 releases up to December 2018. The discography data includes collaborations, featuring appearances, and remixes, yet it lacks information on the popularity of the produced songs. To obtain this additional feature, we combine Discogs with LastFM~\cite{lastfm}, a music providing service that makes play counts of songs available through its API. In addition, we collect genre information of artists from their Wikipedia profiles.


\section{Results}
\label{Sec:results}

\subsection{Dynamics of the top 100 ranking list}
\label{Sec:ranking}

\begin{figure}
\centering
\includegraphics[width=0.90\textwidth]{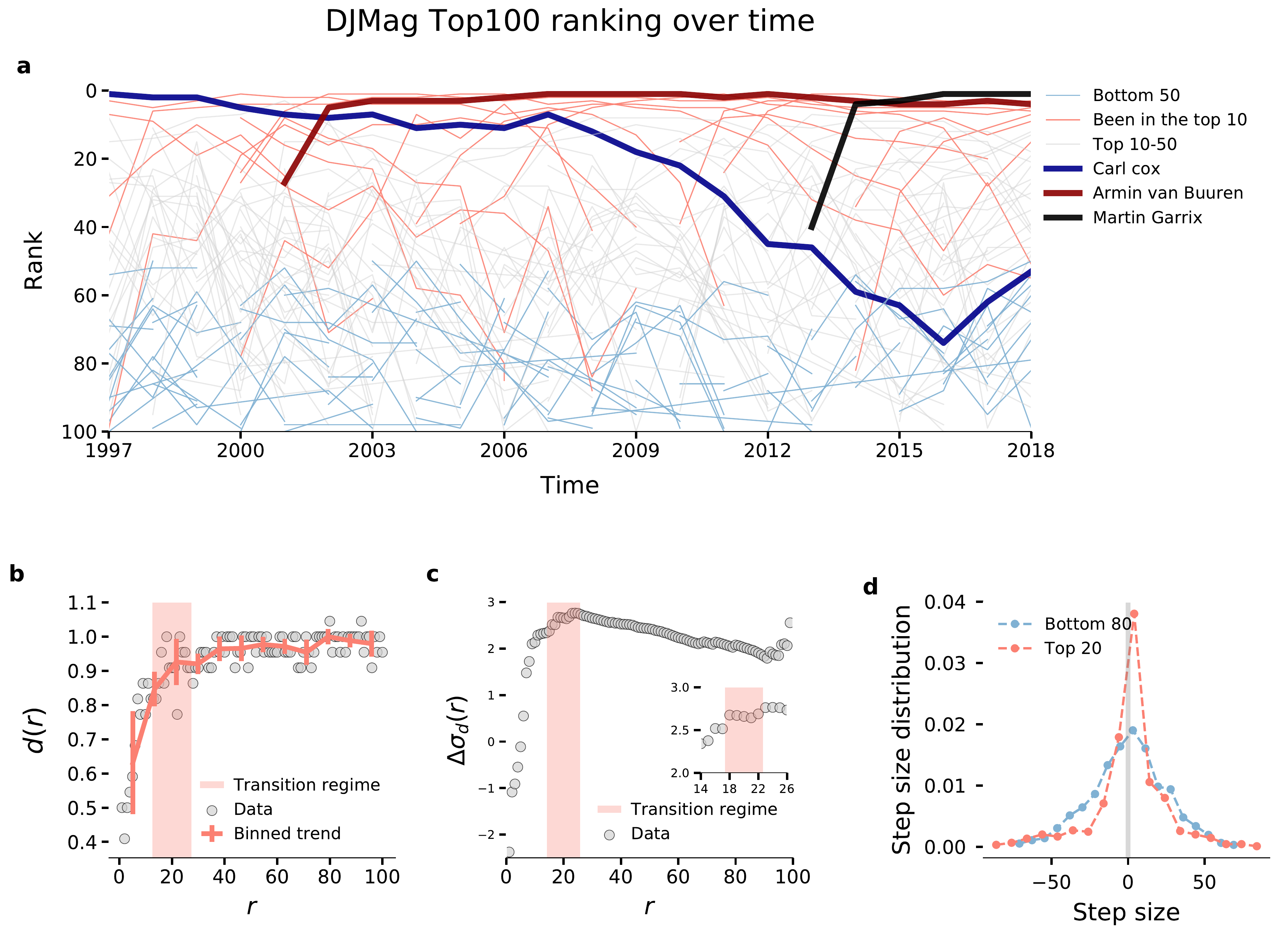}
\caption{ {\bf (a)} Temporal evolution of rank of DJs (denoted by lines) who have made it to the top 100 of DJ Magazine~\cite{djmag}. Colors highlight the part of the ranking list DJs have visited.  {\bf (b)} Rank diversity $d(r)$, defined as the number of individuals that have ever occupied rank $r$ normalized by the length of the observation window.  {\bf (c)} Variance difference $\Delta \sigma (r)$ captured by rank diversity between the real top of the ranking and the rest of the ranking (see Eq. \ref{eq:var}). Inset: Zoomed excerpt of transition regime between boundary of the real top and the rest (axes are the same as in main plot). The width of the transition regime is arbitrary and shown here as 10 ranks before the maximum in variance difference. {\bf (d)} Step size distribution, defined as individuals' rank differences between two consecutive years, comparing top DJs who have ever been in the top 20 (red dots) with the rest (blue dots) (a positive sign means a drop to larger $r$ values, i.e. decline in success).}
\label{fig:fig1}
\end{figure}

During the 22 year-long history of the (public-vote based) ranking list of DJ Magazine, more than five hundred DJs have made it to the top 100. Yet, the electronic music scene has only seen a handful of stars in the ranking list for extended periods of time. Between 1997 and 2018, 11 artists have been crowned as number 1 DJ in the world, a sign of the prominence of figures like Carl Cox,  Tiësto, and Armin van Buuren (ranked among the Top 5 DJs for 17 years). Conversely, success has been ephemeral for most artists: 168 DJs have been in the top 100 only once (with average rank $\langle r \rangle \sim 75.3$), and 99 just made it twice (with average rank $\langle r \rangle \sim72.8$) (Figure \ref{fig:fig1}a).

Such a strong heterogeneity raises the question: What positions in the top 100 can really be associated with well-established success, and where do 'one-hit-wonders' appear? Where is the boundary of being a star DJ, if any? For instance, while DJ Magazine releases names of all top 100 DJs, the top 10 is often treated in a special way, such as being announced name by name at the Amsterdam Music Festival~\cite{amf}. While both 100 and 10 are arbitrary thresholds of success, we manage to identify a threshold that emerges naturally from the dynamics of ranking. To this end, we compute several measures characterizing the ranking list (see Supplementary Information [SI]). We measure the {\it rank diversity} $d(r)$~\cite{cocho2015rank}, which counts the number of different names that appear at a given rank $r$ during the observation period, normalized by the length of this time period ($T$). For instance, 11 different DJs have ever reached the No. 1 position during $T=$22 years, therefore $d(1) = 11/22$.

As Figure \ref{fig:fig1}b shows, a trend-change happens between the upper and lower parts of the ranking. This quantity is somewhat noisy due to limitations on sample size. Therefore, to quantify an actual threshold separating the real top from the rest, we split the ranking into upper and lower tiers based on an initially arbitrary threshold $r$. After that, we compute the variance $\sigma$ of the rank diversity ($d$), and compare lower and upper tiers of the ranking based on the variance difference between them:
\begin{eqnarray}
\Delta\sigma_d(r) = | \sigma_{d, \rho}(\rho \le r) - \sigma_{d,\rho}(\rho > r)|. 
\label{eq:var}
\end{eqnarray}
By computing Equation~\ref{eq:var} for varying arbitrary thresholds (i.e. $r$ values), 
we find monotonously non-decreasing behavior for $\Delta\sigma_d(r)$ in ranks 1--18 and slowly decreasing behavior after rank 22 (Figure \ref{fig:fig1}c) (we neglect the end of the ranking due to noise). Based on the transition between these two regimes (highlighted on the inset of Figure \ref{fig:fig1}c as a maximum in variance difference), we estimate the best splitting boundary separating the top of the ranking from its bottom as $r^* \approx 20 \pm 2$. We refer to the top 20 as the `real' top-tier of the ranking list and call top DJs those who have ever made it to the top 20. For additional measures confirming the stability of the first $r^*$ ranks, see SI section \ref{subsectionSI:realtop}.

Rank diversity $d(r)$ (and further measures in \ref{subsectionSI:realtop}) show a clear difference between high-tier and low-tier DJs, implying that it is more difficult to break into the top 20 than to drop to lower positions of the ranking. As a consequence, once DJs make it to the top 20, they are usually able to maintain their positions with more ease than those at lower ranks (i.e. large $r$). In particular, we find that the yearly rank difference of DJs (Figure \ref{fig:fig1}d) has different trends for top DJs than for those who never make it there: the chances of not changing rank (step size of zero) is twice as high for top DJs than for the rest. 

We find that DJs from the `real' top (as determined from our analysis) have similarly low chances of extremely large rank jumps, despite the fact that from the top (low $r$) there is more rank space to fall down. Yet, unexpectedly large jumps do exist. For instance, a great success of recent years, the American DJ duo The Chainsmokers, started at $r=97$ in 2014 and jumped forward by 79 places to $r=18$, while the Russian trio Swanky Tunes entered the top 100 at $r=97$ in 2015, made a huge jump to $r=27$ the year after, but then fell back to $r=99$ in 2017. To further support these findings we conduct several other comparative measurements in SI section \ref{subsectionSI:realtop2}.

Associating popularity rankings with success in music is not only a potential route to understand the rank dynamics of the most successful individuals, but has a long-standing reputation in the music industry. This way of acknowledging musical success has traditionally used charts and top lists like the DJ Mag top 100 or the well-known Billboard~\cite{billboard}. Another way of capturing the popularity and success of musicians is to measure the number of times people have listened to their songs, e.g., on music providing services~\cite{lastfm, spotify}. We collect both kinds of information about electronic music by combining DJ Magazine's annual top 100 rankings with DJs' song play counts on LastFM~\cite{lastfm}, and measure a low correlation between top DJ ranks and their annual/total play counts on LastFM (see SI section \ref{subsectionSI:correlations}). If raw popularity is not enough, what else do DJs need to reach the top of their profession? In what follows we propose a network-based explanation.


\subsection{Co-release network in the world of electronic music}
\label{Sec:collaboration}

\begin{figure}
\centering
\includegraphics[width=1.0\textwidth]{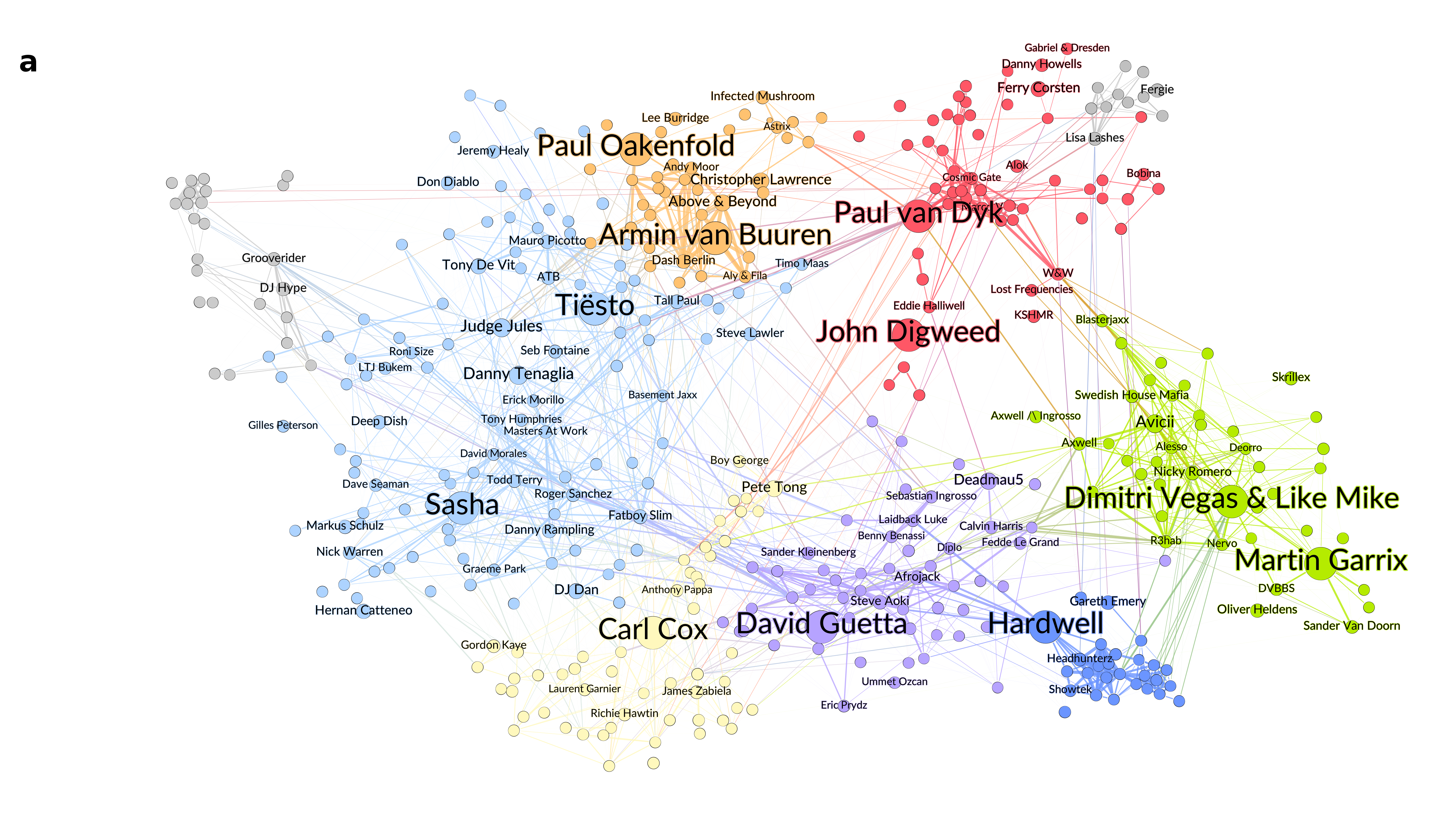}
\includegraphics[width=1.0\textwidth]{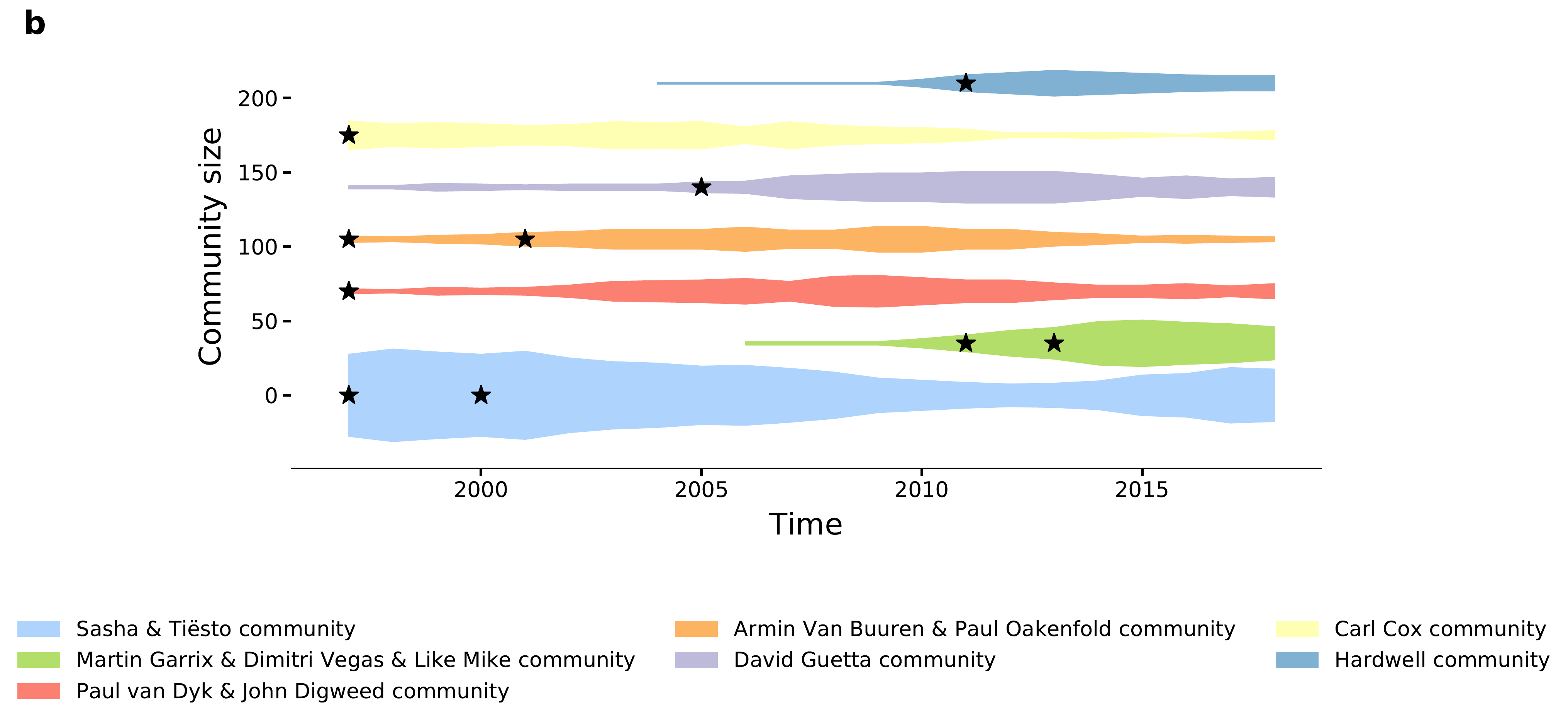}
\caption{{\bf Top 100 DJ network} {\bf (a)} Temporally aggregated and back-bone filtered~\cite{coscia2017network} co-release network of top 100 DJs. DJs are represented by nodes and co-releases by links between them, with link width proportional to the number of releases DJs collaborated on. Node size is proportional to the DJ's best rank (larger size means lower $r$ and higher success). Node colors show the detected music communities~\cite{blondel2008fast}. Top 20 DJs are shown by name.  {\bf (b)} Temporal growth of DJ communities, with size measured as the number of DJs in a given year's top 100 ranking. Black stars denote the entry years of the named (later No. 1) DJs.}
\label{fig:fig2}
\end{figure}

Are there any network effects that keep a handful of stars at the top, and result in a faster dynamics of rank change at the bottom?  What is the relationship between music collaborations and the observed dynamics of the top 100 DJ ranking list?  To address these questions, we construct and analyze the co-release network of top 100 DJs based on their profiles as electronic music artists on Discogs~\cite{discogs}. We find that having a central position in the co-release network correlates with success (SI Section \ref{subsectionSI:nwcorrel}); however, after visualizing the network, we find that it does not have a single central region of stars but shows a non-trivial modularity structure (Figure \ref{fig:fig2}a). For further analysis, we extract the back-bone structure of the network by using the recently introduced noise-corrected filter method~\cite{coscia2017network} (for details see SI section \ref{subsectionSI:netviz}). The analysis of the community structure of this back-boned network, determined by a widely-used simple heuristic method~\cite{blondel2008fast}, reveals seven communities covering 92\% of the nodes in the top DJ network. Surprisingly, each community includes one or two DJs who once earned the No. 1. DJ title.

We study the temporal evolution of DJ communities by measuring their size, defined as the number of top 100 DJs in each group over time.  We find that the communities, named after their leading artists, rise and fall over time distinctively (Figure \ref{fig:fig2}b), highlighting how new artist communities form and old ones fade away. These temporal trends are in agreement with recent findings on changes in fashion cycles and the roles of the elite in  them~\cite{klimek2019fashion}. We see that the mainstream of electronic music emerges in the community of Sasha and Tiësto, and how the latest electronic dance music trends grow around Dimitri Vegas \& Like Mike and Martin Garrix. We find a significant correlation ($\approx0.73$ on average) between community size and the average rank of the three most successful artists in the community.  This observation further supports the major role of leading artists in the growth of their community and even music scenes themselves.
By studying the entry time distributions of DJs in the various communities (see SI section \ref{subsectionSI:commtimescales}), we find that these leading figures typically enter their communities amongst the earliest members (Figure \ref{fig:fig2}b). In other words, leading artists are not simply the most popular, but also some of the founding members of their communities (further details of these differences in SI sections \ref{subsectionSI:comms}-\ref{subsectionSI:commtimescales}). 

Uncovering the community structure of the top DJ music scene and observing its distinct temporal trends lead us to ask the question: what are the main differences between these communities? One possibility is that musical genres (such as techno, house, and trance) reflect these differences.  To test this hypothesis, we collect genre information on the top DJs from Wikipedia. Out of the 420 artists present in the network, 251 have genre information, from a pool of 64 subgenres of electronic music, with 3.2 tags per DJ on average. We reconstruct the genre-tag distribution of each community, characterized by the genre vector ${\bf g}_i$ for community $i$ such that  ${\bf g}_{i,j}$ equals the number of DJs in community $i$ that are associated with genre tag $j$. In this way, we compute the genre-similarity $\Gamma$ of two communities, $l$ and $m$, as the cosine-similarity of their genre vectors:
\begin{eqnarray}
\Gamma_{l,m} = \frac{{\bf g}_l \cdot {\bf g}_m}{|{\bf g}_l||{\bf g}_m|}.
\end{eqnarray}
We find that the major genres in the newly emerging communities are usually moderately different, with an average cosine similarity of $\overline{\Gamma} \approx 0.445$, in agreement with recent results on changes in fashion trends~\cite{klimek2019fashion}.

We conclude that the two most alike communities are both focused on trance and progressive, have a similarity score of $\Gamma \approx  0.84$, and are led by Paul van Dyk and Armin van Buuren. We also find that these two communities are the closest in time, with average debut years of 2005 and 2006. In contrast, the two most different communities are led by Martin Garrix (joined in 2013) and Carl Cox (joined in 1997), with a similarity score of only $\Gamma \approx  0.14$ and with more than a decade difference in typical debut years. While DJs in the former group are mostly playing house music, the latter is more focused on techno. As these time differences already suggest, the further two communities peak from each other (at time $t_{p,i}$ for community $i$), the more different their genre profiles (${\bf g}_i$) are. We show this effect by computing the time difference between peak years of the pair of communities $l$ and $m$,
\begin{eqnarray}
\tau_{l,m} &=& |t_{p,l} - t_{p,m}|,
\end{eqnarray}
and correlate those values with the genre-similarity score $\Gamma_{l,m}$. We get a Spearman correlation of $r_{\Gamma \tau} \approx 0.62$, supporting our claim that the closer two communities peak in time, the more similar their genre distribution is. More details on the similarities of communities are in SI section \ref{subsectionSI:similarity}. The main trends of these genre differences, illustrated by genre tags, are summarized in Table \ref{tab:tab1}: while in the late '90s and early 2000s house and techno were the most popular genres, by the middle of the 2000's trance and progressive house started gaining popularity, mostly driven by Armin van Burren, who has been in the top 5 ever since.

Overall, we report that the top 100 DJs form different, temporarily separable communities, and these communities represent slight changes in musical trends. Each community has typically one or two leading figures, who are one of the first and typically the most successful members of their communities. These observations suggest that top, central DJs act as gatekeepers by constantly renewing the field of electronic music, and they shape both music trends and communities by bringing in new artists. In what follows we further explore the existence of such a mentorship effect.

\begin{table}
\scriptsize
\begin{tabular}{lllll}
\textbf{Community - Lead DJs (debut year)} & \textbf{Genre 1} & \textbf{Genre 2}   & \textbf{Genre 3}      & \textbf{Average debut year}             \\ \hline \hline
 Sasha (1997), Tiësto  (2000)              & house                 & electronica             & techno             & 2000                                                  \\   \hline
Carl Cox   (1997)                                   & house                 & techno                     & electronica        & 2002                                               \\    \hline
Armin van Buuren (2001),                  & trance                & progressive house & electronica        & 2005                           \\
 Paul Oakenfold (1997)                        &                            &                                                                    &                             &     \\    \hline
Paul van Dyk (1997) ,                           & trance                & progressive house & progressive trance & 2006                                          \\
 John Digweed (1997)                           &                            &                                                                    &                             &     \\    \hline
David Guetta  (2005)                            & house                 & electro house     & progressive house  & 2008                                                       \\ \hline
Hardwell        (2011)                             & hardstyle           & progressive house & dutch house        & 2012                                \\   \hline
Dimitri Vegas \& Like Mike (2011),    & electro house   & progressive house & big room house     & 2013                 \\
 Martin Garrix  (2013)                          &                             &                                                                    &                                &      \\
\end{tabular}
\caption{{\bf Genre distributions in DJ communities.} Name and debut year of the No. 1. DJs of each community, the three most frequent genres of the DJ communities, and the average debut year of artists in each group.}  \label{tab:tab1}        
\end{table}


\subsection{Mentorship in electronic music}
\label{Sec:mentorship}

Our results show that most communities in the electronic music scene contain one or two No. 1 DJs who make an appearance in the early stages of each community's life-cycle. How do these groups form? What are the major social forces shaping the DJ world? Do newcomers join existing groups independently, or are they more likely to be brought in by their former collaborators? In other words, does collaborating with top 100 DJs help new artists make it to the ranking list and join the music elite?

\begin{figure}
\centering
\includegraphics[width=0.9\textwidth]{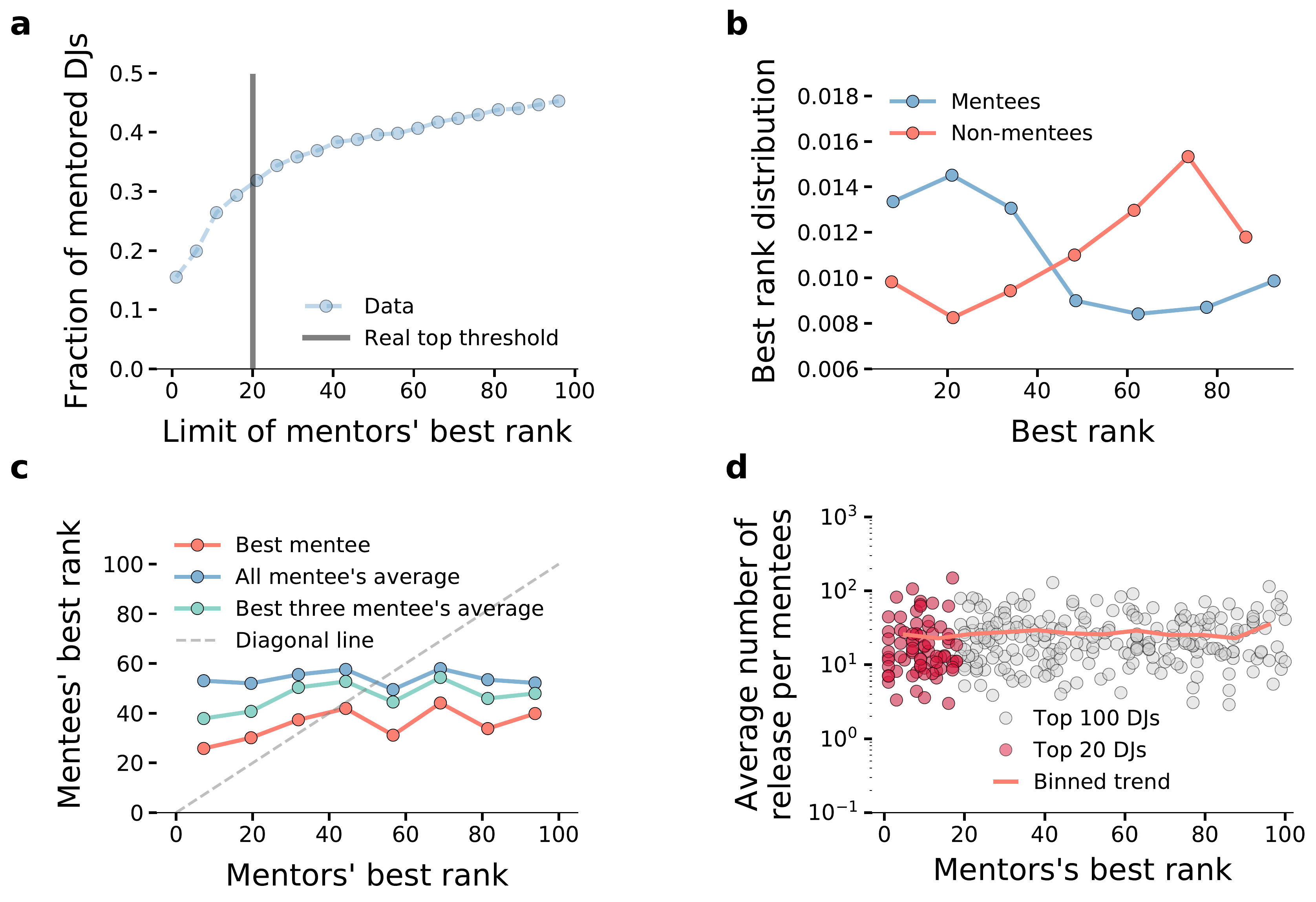}
\caption{{\bf  Mentorship in electronic music} {\bf (a)} Fraction of mentee DJs who have been mentored by artists with a best rank not lower than the limit rank measured on the horizontal axis. The vertical line represents the threshold of the top 20, who mentored more than $\sim0.3$ of all mentored DJs. {\bf (b)} Comparison between the (percentile-binned) distribution of the best rank of the DJs who were mentored (blue line) and those who were not (red line). {\bf (c)} Average best rank of mentees as a function of mentors' best rank. Mentees correspond to three groups: the best mentee of a mentor (red continuous line, Spearman rank correlation $r_s\approx0.199$), avereage best rank of its the best three mentees (green continuous line, $r_s\approx0.192$), and average best rank of all its mentees (blue continuous line, $r_s\approx0.035$). The diagonal line illustrates an ideal case where mentees reach similar best ranks as their mentors. {\bf (d)} Number of releases normalized by number of mentees for mentor DJs, expressing the frequency of their mentoring activites, and measured as a function of the mentor's best rank. Top 20 DJs are highlighted by red (Spearman rank correlation $r_s\approx0.04$) and the rest by grey ($r_s\approx-0.12$).}
\label{fig:fig3}
\end{figure}

Known success stories and anecdotes, like Rolling Stone magazine's take on Afrojack and David Guetta~\cite{rollingstone}, suggest that mentoring plays an important role. To investigate this hypothesis, we define {\it mentorship} (a special type of networking behavior~\cite{higgins2001reconceptualizing,sekara2018chaperone}) among top DJs in the following way: DJ$_1$ is the mentor of DJ$_2$ if they both made it to the top 100 ranking respectively at times $t_1$ and $t_2$ (with $t_1< t_2$) and if they first appeared on the same release earlier than $t_2$. We find that about half of the DJs that ever made it to the top 100 have been mentored before, and about 30\% of them were mentored by DJs with a best rank of 20 or better (Figure \ref{fig:fig3}a), implying that the role of the most successful individuals is central in community building, by means of the mentoring of new artists.

Our results suggest that the most successful DJs build communities around themselves. Is this beneficial only for them, or does it also boost the expected success of their mentees? To answer this question, we compare the distribution of the best rank of top 100 DJs, differentiating between DJs that have been mentored before or not (Figure \ref{fig:fig3}b). Mentee DJs have a significantly higher chance of achieving ranks in the top, and a large fraction of them even approaches the edge of the top 20. On the other hand, DJs who have not been mentored typically just show up at the tail of the top 100 and have negligible chances of making it to the top 20.

One side of the formula is clear: mentorship boosts the expected success of newcomers, which aligns with previous findings on mentoring in science~\cite{malmgren2010role,sekara2018chaperone}. However,  we also see a clear boundary between non-successful DJs and all-time stars, which makes us wonder: are star DJs star mentors as well? We tackle this problem by comparing the average best rank of mentees to the best rank of their mentors. We find that mentees only profit slightly from having high-profile mentors, since the mentees' expected best rank barely improves for highly successful mentors. This is also captured by the low correlation between the best rank of the mentors and the average best rank of their most successful mentees (Figure \ref{fig:fig3}c). In other words, no matter how successful a mentor is, the expected success of their mentees is capped and is slightly below the real top, even for the best mentees. Moreover, if we compare the number of mentees each DJ has, relative to the number of releases they produce, we measure a low correlation (for top 20 DJs $r_s \approx 0.04$, see Figure \ref{fig:fig3}d). The fact that most mentees are mentored by top DJs is thus simply due to top DJs being more productive. Therefore, star DJs do not carry an extra 'star-mentor' effect; all DJs seem to follow the same pattern and simply release more music when they collaborate more, which includes co-releases with new artists. A cumulative advantage process may help top DJs keep their top positions, since the more successful DJs are, the more resources they have access to, which leads to higher chances of recruiting new mentees, as well as new and even more releases.

Taken together, our results imply that mentorship plays an important role in the rise of new stars and the growth of their prolific environment, but mentorship alone is not enough to explain the emergence of superstars. Such events seem to depend on (as of now) unknown mechanisms that cannot be inferred solely by an analysis of the music co-release network.

\section{Discussion}
\label{Sec:discussion}

Electronic music, as one of the most popular music genres, has evolved into a complex ecosystem, with DJs and producers releasing and collaborating together across multiple subgenre styles over the past two decades. Here we have investigated the temporal evolution of this field, focusing on how to pinpoint and distinguish a longstanding elite from the rest of electronic musicians. We have also proposed potential mechanisms that could lead to the differences between elite musicians and less successful artists. First, we connect the dynamics of the top 100 ranking list of electronic music artists to their underlying co-release patterns, in order to infer major principles of success. We find that the historical top 100 rank splits into two distinct regimes in terms of the stability of their dynamics, showing the existence of a persistent elite in the DJ world. From collaboration patterns, we show that those superstars who have reached the No. 1 position usually tend to lead segregated communities, which rise, peak, and fall separately over time, often representing changes in genres. We also see that a major social force driving these communities is mentorship, since new DJs usually join the top 100 after co-releasing albums with already established artists. DJs who have been mentored before seem to perform significantly better, yet even their chances of overcoming their mentors are slim. We report that while star DJs exist, star mentors do not: the success of mentors has little influence on the expected success of their mentees.

While our results highlight interesting and major patterns in the growing ecosystem of electronic music artists, they have some limitations. The top 100 ranking of DJs reflects the opinion of a particular segment of electronic music fans, mostly limited to online platforms. Live shows and festivals, also a major platform of electronic music, are disregarded. This shortcoming may be alleviated by incorporating data from social media and other music providing platforms (to have a less biased picture of the online landscape), or by using information about live shows, tickets and record sales to connect our work with offline behavior. Another major question is how well our findings generalize to other genres. Are the observed phenomena particular to electronic music, or do rock, pop, and other musical genres follow similar trends? Since various rankings exist for other genres, such as in Billboard Magazin~\cite{billboard}, and collaboration and co-release data are also available (for instance on Discogs), most of our analysis is replicable and may be tested in the near future.

Our results suggest that the realm of electronic music is driven by a long-standing elite, which substantially boosts the expected success of unknown artists via mentorship. And yet, the same elite seems to be systematically preventing outsider artists from joining it. Further characterization of these elite star DJs may require other approaches and data sources, such as metadata on musicians. In this way, we could analyze and incorporate biases based on, e.g., gender and birth location. It is also possible that our dataset, limited to releases, does not capture more complex and social levels of mentorship, such as earlier interactions (i.e. the first time junior artists meet future mentors or even their managers, potentially much earlier than a first co-release).

Possible venues of related future research include an understanding of the differences between the trajectories of those we never make it to the top 100 against those who do and the analysis of the early-career patterns of these two groups. As a step further from descriptive analysis, an interesting direction is the development of predictive models that capture not only the next top 100 or No. 1 DJ's identities, but also the next new entries: people who are already out there with the potential for becoming the stars of the next generation. In this direction, we would also suggest to study musical features and extract various descriptors of the audio data itself, as well as combine the collaboration network with co-follow networks extracted from various social media outlets. 

We propose here a first attempt to understand the emergence of success in electronic music by obtaining quantitative findings on the existence and behavior of an exclusive elite of star DJs and producers. These results not only give insights into an interesting and wildly dynamic social system, but also offer a good starting point for further research and policy suggestions. These include directions such as how to make electronic music more inclusive and less biased, help junior artists to be less exposed to long-standing stars, and make steps towards more merit- (and less business-based) spaces for artistic creativity.

\section{Data accessibility}
Datasets and code to reproduce the findings that lead to the main conclusions of the paper can be found at: \url{https://github.com/milanjanosov/Elites-communities-and-the-limited-benefits-of-mentorship-in-electronic-music}

\section{Authors' contributions}
All authors designed the study and contributed to the manuscript. M.J. performed the data collection, analysis, and numerical results.

\section{Competing interests}
The authors declare no competing financial interests.

\section{Acknowledgement}
The authors wish to thank Dávid Deritei and Manran Zhu for their helpful suggestions. The project was partially supported by the European Cooperation in Science \& Technology (COST) Action CA15109.

\newpage

\renewcommand\thesection{S\arabic{section}}
\renewcommand\thefigure{S\arabic{figure}}
\renewcommand\thetable{S\arabic{table}}

{\bf \huge Supplementary Information for \\ \hspace{1cm} } 

 { \Large \bf Elites, communities and the limited benefits of mentorship in electronic music}


\section{Dynamics of the top 100 ranking list}


\subsection{Is the top 100 the real top?}
\label{subsectionSI:realtop}

\begin{figure}[htp]
\centering
\includegraphics[width=1.00\textwidth]{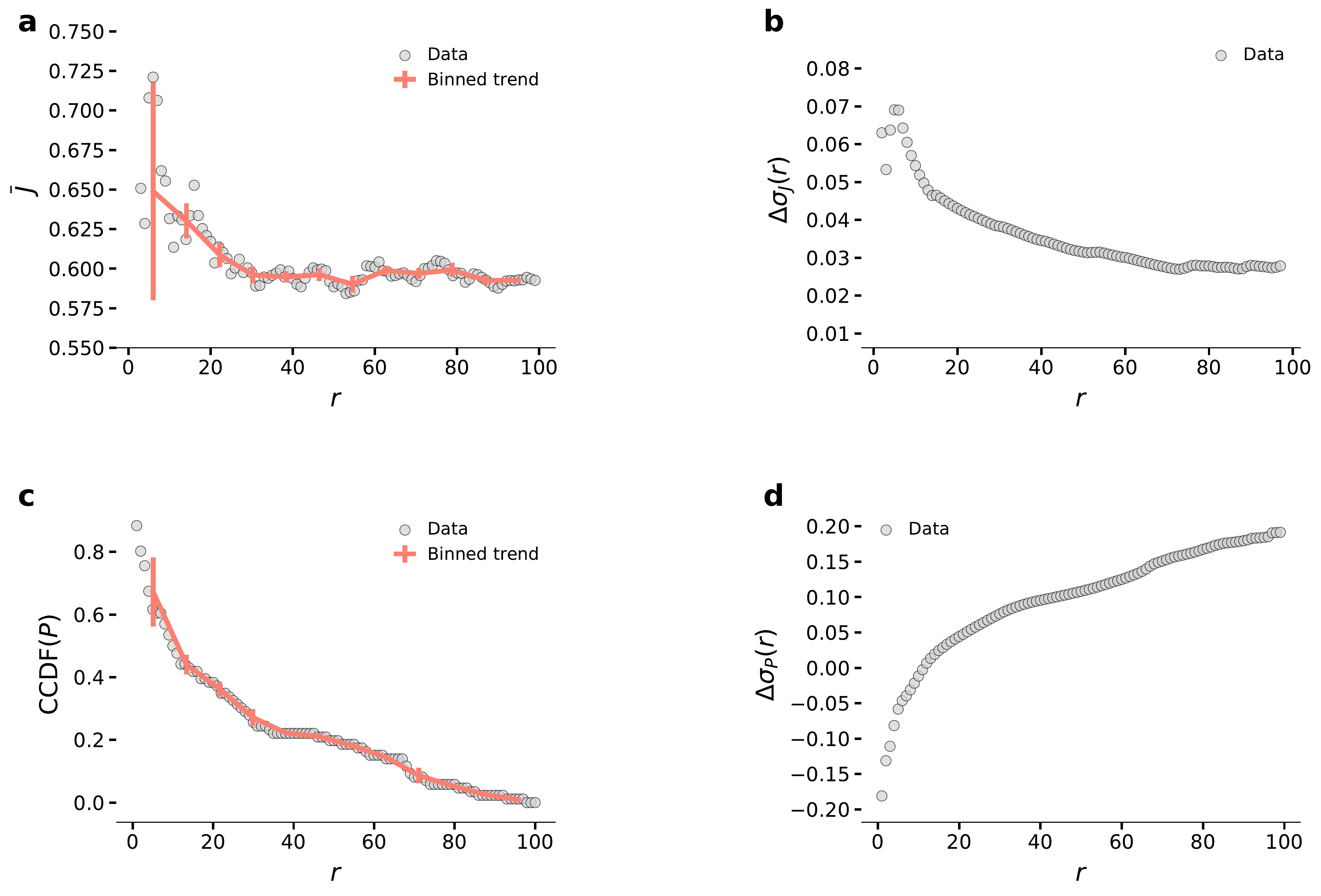}
\caption{{\bf Homogeneity of the ranking.} The ({\bf a-b}) subfigures show the Jaccard similarity $\overline{J}$ and its splitting variance based on Eq. \ref{eq:var} in the main text, while ({\bf c-d}) visualize the CCDF of the  remaining probability $P$ as a function of the splitting rank $r$ and its splitting variance, further pointing out the differences between the top and the bottom of the ranking.} 
\label{fig:topbottomcomp}
\end{figure}

In addition to the study of rank diversity in Section \ref{Sec:ranking}, we define and compare the following two measures of the ranking list in order to show the difference between the top and the bottom of the ranking (results are also shown on Figure \ref{fig:topbottomcomp}):

\paragraph{Jaccard similarity across years.} We compare set of DJs being present at rank $r$ at a given year $t$ ($P_r(t)$) to the DJs in $r$ at time $t+1$  ($P_r(t+1)$), average this over time, and study it as a function of $r$ (Figure \ref{fig:topbottomcomp}a):
\begin{eqnarray}
\overline{J_r} & = &    \bigg\langle \  \frac{P_r(t) \cap P_r(t+1)}{P_r(t) \cup P_r(t+1)} \bigg\rangle_t .
\end{eqnarray}
By analyzing the behavior of this measure as a function of $r$ (Figure \ref{fig:topbottomcomp}a,c), we see a regime change from a monotounsly decreasing trend to a flat line around $r^* \approx 20-25$, which overlaps with our finding presented in Section \ref{Sec:ranking}. The same trends are also visible on the variance differences if we split the ranking at $r$ according to Eq. \ref{eq:var}  in the main text (Figure \ref{fig:topbottomcomp}b,d).

\paragraph{Remaining probability.} The probability $P$ of an individual taking the same rank position at time $t$ and time $t+1$. For a certain rank $r$ we compute the number of times the same individual occupies that rank at time $t$ and $t+1$,  and average it over time. Figure \ref{fig:topbottomcomp}c shows the complementary cumulative distribution function (CCDF) of this probability as a function of $r$ and shows that there is a clear change in trend around the 20th position, further supporting that neither top 10 nor top 100 are special; however, somewhere in-between there is a clear separation between a stable, popular elite, and the rest of musicians. We also measure the splitting variance difference (Eq. \ref{eq:var} in the main text), which shows a change in trend around $r^*\approx 20$ as well (Figure \ref{fig:topbottomcomp}d).

\subsection{How does the real top differ from the rest?}
\label{subsectionSI:realtop2}

\begin{figure}[htp]
\centering
\includegraphics[width=0.9\textwidth]{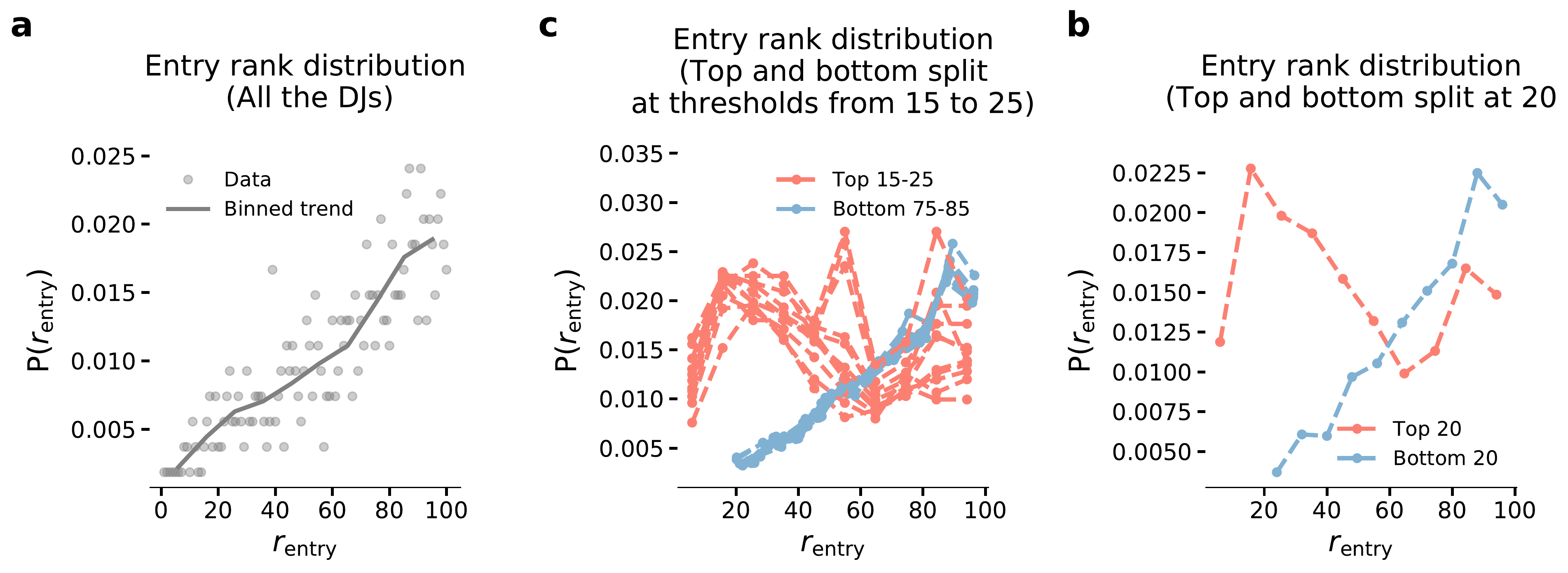}
\caption{{\bf Entry rank distribution.} ({\bf a}) The entry rank distribution of the total population. ({\bf b}) The entry rank distribution of the top and bottom DJs, split based on their best rank with different splitting thresholds (from 15 to 25). ({\bf c}) The entry rank distribution of the real top DJs and the rest of the ranking.}
\label{fig:entry}
\end{figure}

Here we compare the distribution (Figure \ref{fig:entry}a) of the entry rank of the DJs over the entire population, and at various splitting thresholds, including splitting between the identified real-top boundary (Figure \ref{fig:entry}b-c). As for the distribution of entry ranks, we see that there are increasingly more people entering the ranking at lower ranks. However, if we split individuals at the threshold of the real top, we see noisy but quite different trends. Having said that, the trends seem to be inverted: while `bottom' individuals are increasingly more likely to enter at lower ranks, high achievers are more likely to enter at higher places immediately.


\subsection{Relationship between the best ranks and the songs' raw popularity}
\label{subsectionSI:correlations}

\begin{table}[htp]
\scriptsize
\begin{tabular}{l|llllll}
                                     				                                             & December 20    & February 17  & March  22    & Apr 19 &  May 26 \\  \hline
Total play count of songs in 2018		                                             & 14,669,041              & 17,870,288          & 27,181,518      &  17,831,911   & 18,007,282\\
Corr(Rank last year vs. play count of songs released in 2018)        & 0.394                      & 0.393                   & 0.371                & 0.346    & 0.35 \\
Corr(Best rank ever vs. total play count over the career)                   & 0.262                     & 0.321                    & 0.332               & 0.318  & 0.319
\end{tabular}
\caption{{\bf Play count and ranking correlations.} Correlation between number of songs (released in 2018--2017) and the ranking of 2018; play counts of songs released during several 2-3 day-long crawling periods (marked by their start dates); and correlation between the best rank of DJs and the overall play count of their songs. The decreasing trend of total play counts shows how songs released in 2018 are losing popularity in 2019.} \label{tab:tab1}
\end{table}

\begin{figure}[htp]
\centering
\includegraphics[width=1.0\textwidth]{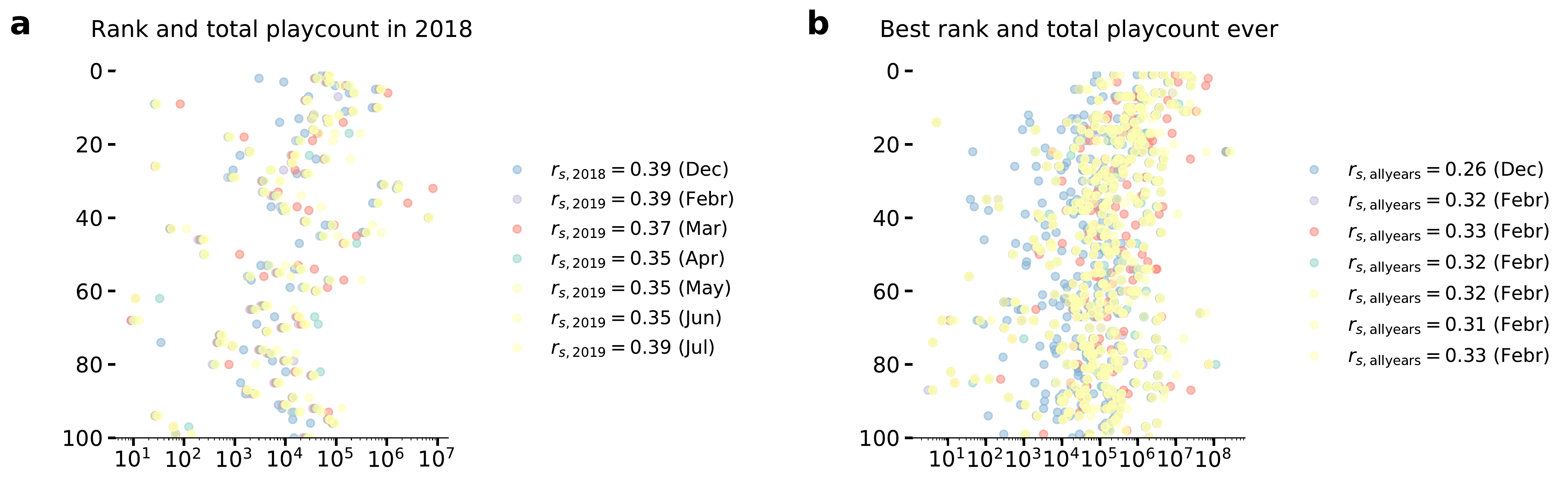}
\caption{{\bf The absolute values of the computed Spearman-rank correlations between songs' popularity and top 100 ranking. } ({\bf a}) Correlation plot between the ranks of DJs (in 2018) and the total play count of their songs released during the year before this ranking. ({\bf b}) Correlation between the total play count of the songs ever released by these DJs, and their best rank on the top 100 ranking ever.} 
\label{fig:corr}
\end{figure}

The top 100 ranking is based on the number of votes coming from the fans of DJs. The raw vote counts and ranks below 100 are not available. We compare the results of last year (2018) in the following way. We consider those DJs who were present in the top 100 in 2018 (announced on 20 October 2018). Then we pick all the songs of these DJs that were released before the ranking of 2018 came out, but after the ranking of 2017 came out, covering a period of one year. Then we compute the overall popularity (total play count) of these songs at different points in time (as detailed in Figure \ref{fig:corr}) and compute their correlations with the overall ranking. We also correlate the best rank ever achieved by the top DJs to the total play count of the songs they released (Figure \ref{fig:corr}b), where we find an even lower correlation. The correlation values are summarized in Figure \ref{fig:corr}) and Table \ref{tab:tab1}. 

\pagebreak
\newpage
\section{Co-release network in the world of electronic music}

\subsection{Network visualization}
\label{subsectionSI:netviz}

To obtain the network visualization on Figure \ref{fig:fig2}a we go through the following steps. First, we construct the (quite dense) original sub-network of top 100 DJs, with 15,403 edges distributed among 486 nodes in the giant component. Then we apply network filtering algorithms before further analysis. We use the recently introduced noise-corrected filter method~\cite{coscia2017network}, with which we filter out $\sim88\%$ of edges while keeping $\sim86\%$ of nodes. To extract communities we use an established heuristic method~\cite{blondel2008fast}.

\subsection{Network centralities and success}
\label{subsectionSI:nwcorrel}

\begin{figure}
\centering
\includegraphics[width=0.7\textwidth]{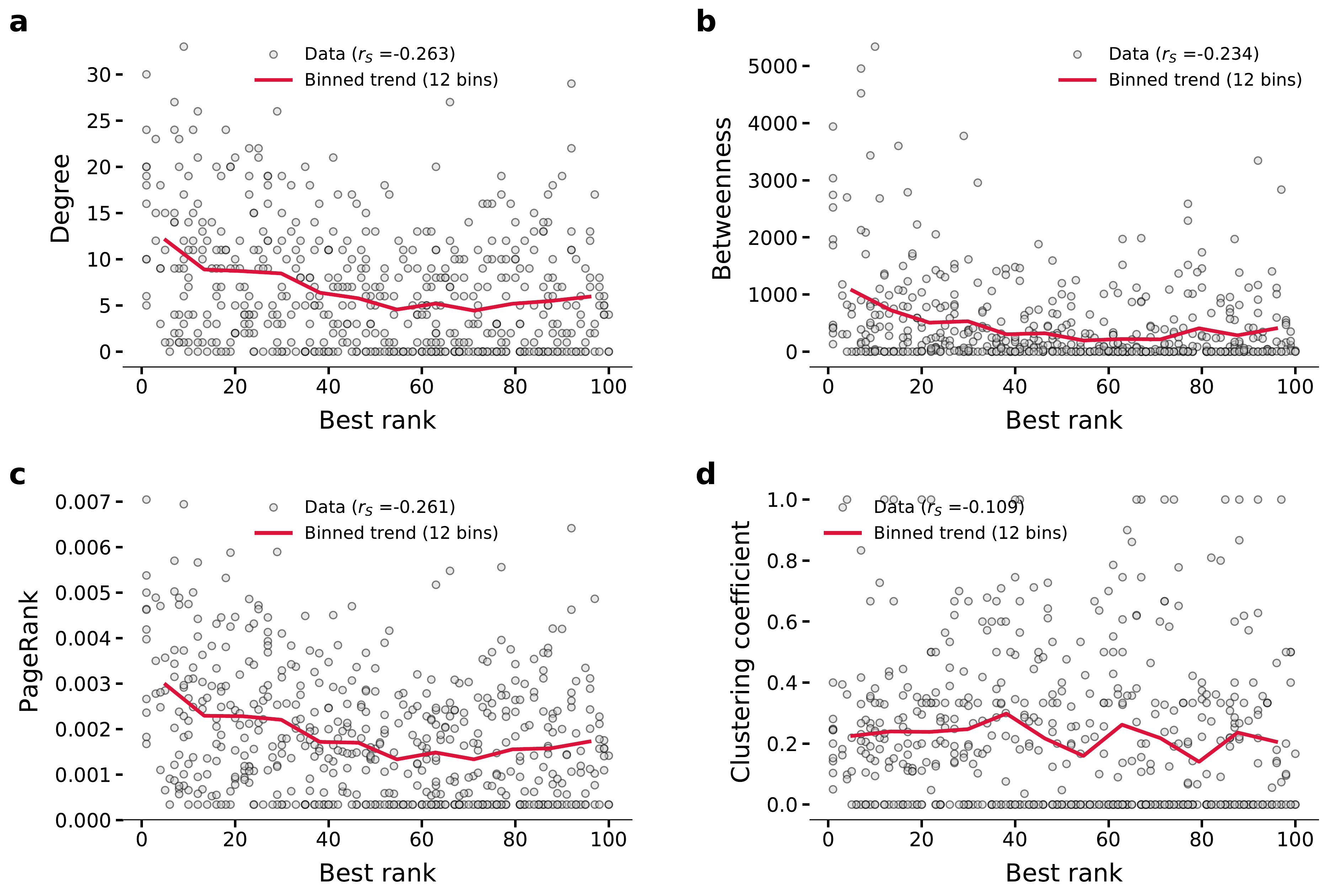}
\caption{{\bf Correlations between the best rank and network centralities.} The red line visualizes the binned trends of the centralities (node degree, betweenness centrality, PageRank centrality, and clustering), while individual points represent individual DJs.}
\label{fig:networkbestrank}
\centering
\includegraphics[width=0.7\textwidth]{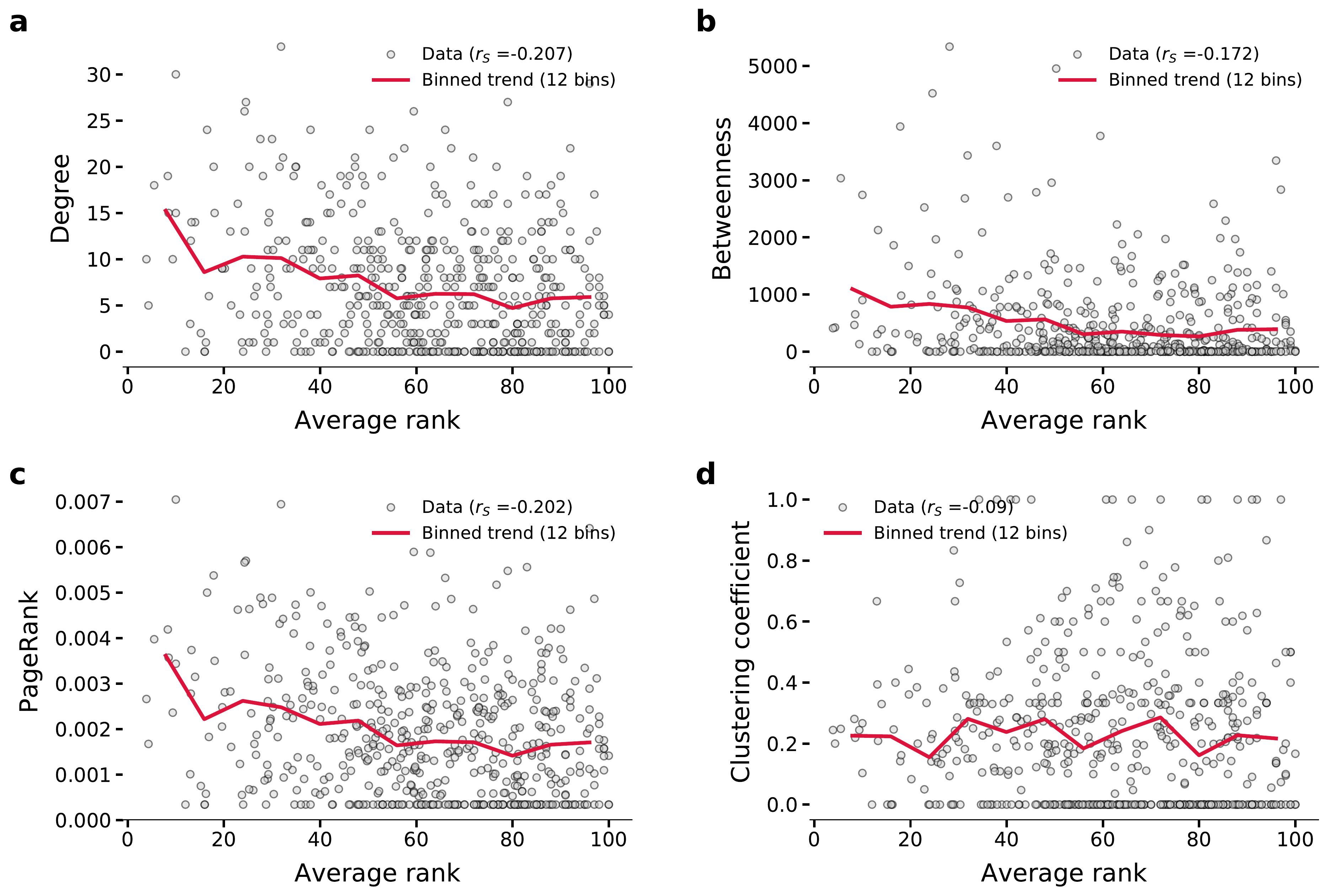}
\caption{{\bf Correlations between the average rank and the network centralities.} The red line visualizes binned trends, while individual points represent individual DJs.}
\label{fig:networkavgrank}
\end{figure}

Here we show the correlations between the best and average ranks that the top 100 DJs have achieved and their measured network centralities (captured by degree, betweenness, and PageRank centrality), as well as the clustering coefficient. The results, highlighting a surprisingly low correlation are shown in Figure \ref{fig:networkbestrank}.\ref{fig:networkavgrank} and summarized in Table \ref{tab:SItab2}. 


\begin{table}[b]
\scriptsize
\begin{tabular}{lllll}
             & Degree & Betweenness & PageRank & Clustering \\
Best rank    & 0.263  & 0.234       & 0.261    & 0.109      \\
Average rank & 0.207  & 0.172       & 0.202    & 0.09     
\end{tabular}
\caption{The correlation between the different centrality measures and the best and the average ranks of the DJs.}
\label{tab:SItab2}
\end{table}


\subsection{Communities over time}
\label{subsectionSI:comms}

\begin{figure}[htp]
\centering
\includegraphics[width=0.8\textwidth]{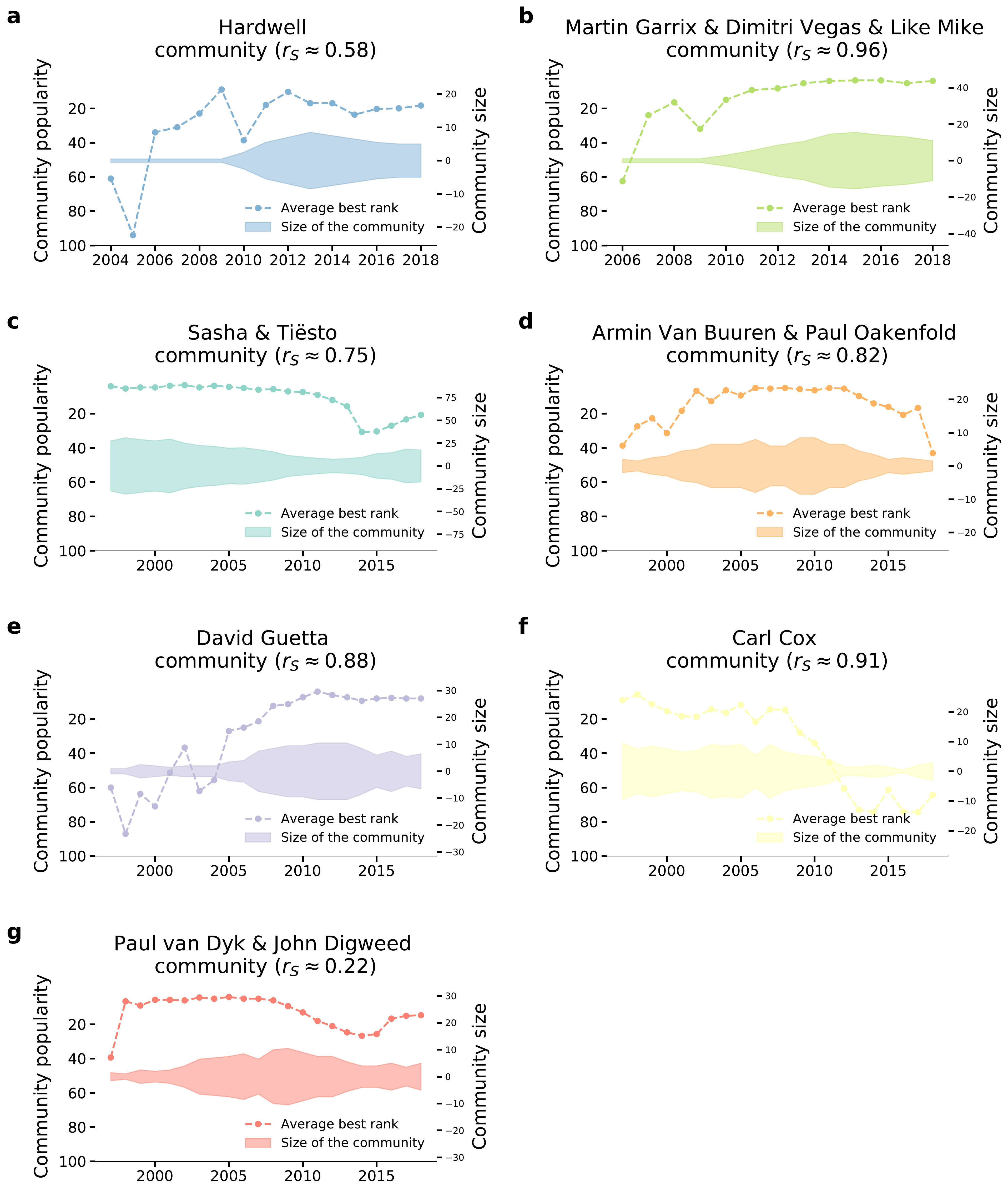}
\caption{{\bf  Temporal dynamics of communities in electronic music.} Average popularity of the three highest-ranked DJs of each community (colored dashed lines), and size of the community (shaded area over time). Titles include the Spearman rank correlation $r_s$ between the two quantities over time. } 
\label{fig:SFigure5.png}
\end{figure}

On Figure  \ref{fig:SFigure5.png} we show how the size of DJ communities changes over time, and how it correlates with the popularity of the communities over time. We quantify popularity by taking the average rank of the three highest ranked DJs of each of the seven largest communities of the giant component of the top 100 DJ's network. The figures show high correlation, implying that the more popular a community is, the more members it has in the top 100.

\subsection{Typical timescales of DJ communities}
\label{subsectionSI:commtimescales}

\begin{figure}
\centering
\includegraphics[width=1.0\textwidth]{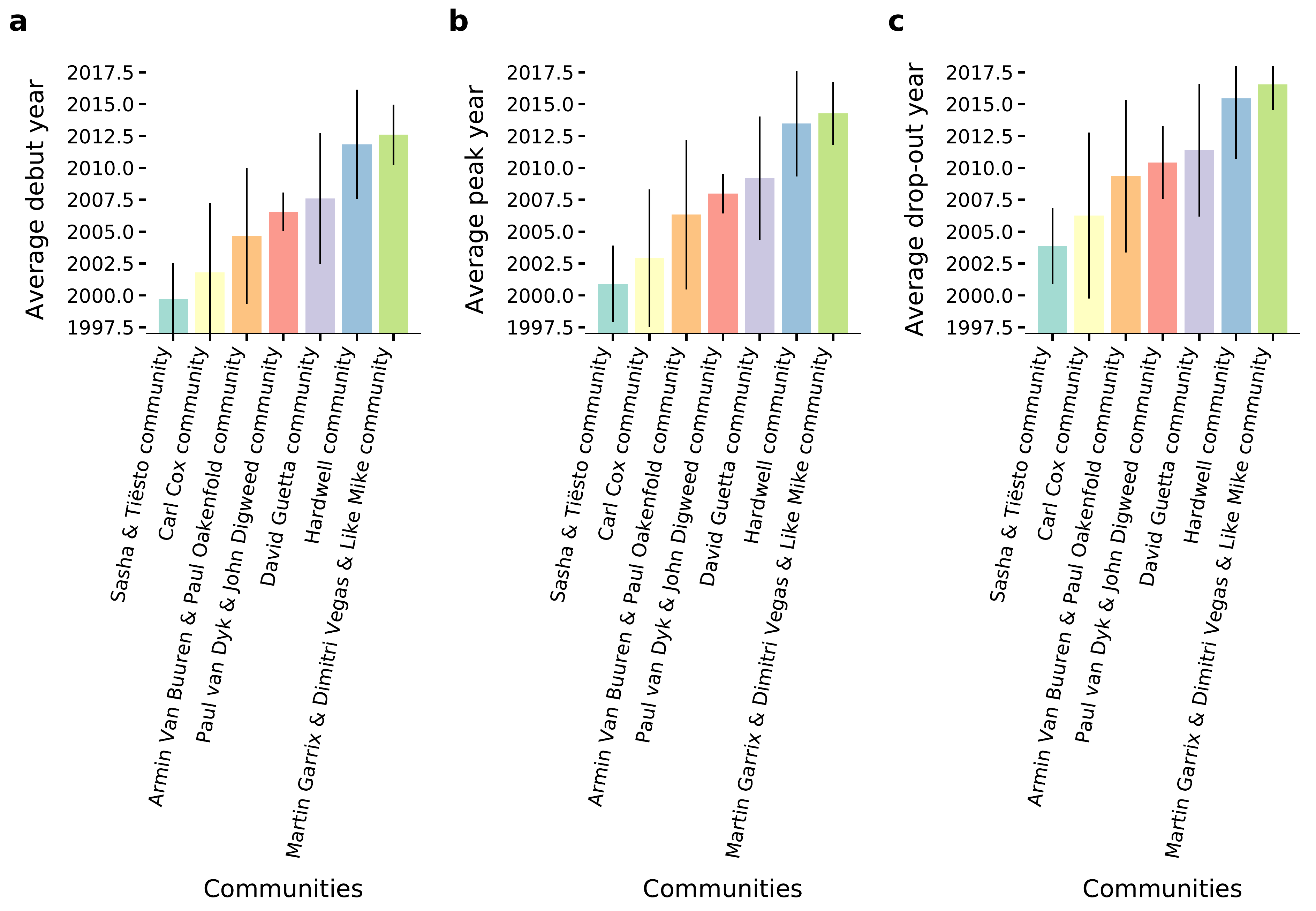}
\caption{{\bf Average time scales of DJ communities.} The average debut year, peak year (reaching the highest average rank), and drop-out years for the detected communities, different colors denoting the different communities.}
\label{fig:avgtime}
\end{figure}

\begin{figure}
\centering
\includegraphics[width=1.0\textwidth]{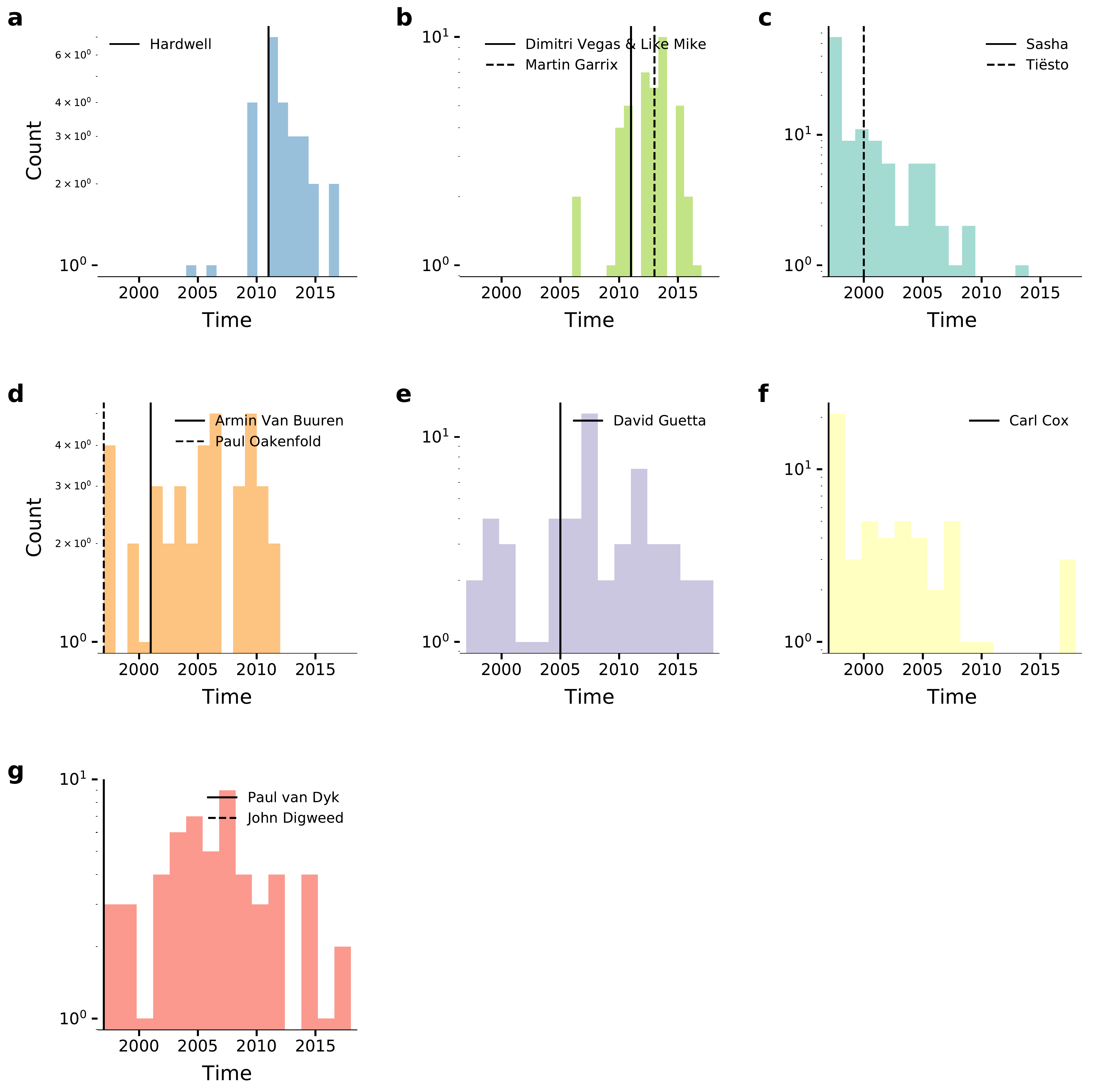}
\caption{{\bf Entry rank distributions of DJ communities.} Distributions of entry years of DJs in each community, with vertical lines for the entry years of each community's leading figures highlighting their early arrivals.}
\label{fig:entryrankcomm}
\end{figure}

We analyze the typical timescales of the different communities: the typical entry, drop-out, and peak years of the DJs belonging to all communities. As visualized in Figure \ref{fig:avgtime}, there is a clear temporal order between different communities. We also compare the distribution of entry years of DJs across communities (Figure \ref{fig:entryrankcomm}), where we see that the community-leading figures have typically entered the ranking at the early stages of their communities' lifetime.

\subsection{Genre-similarities of DJ communities}
\label{subsectionSI:similarity}

In each DJ community a set of genres is present with varying degree. We encode this in the genre vector ${\bf g}_i$ for community $i$, were each element corresponds to a genre that was observed in the data set, and is equal to the number of times that genre occurred through the member artist profiles. For instance, if there are only two genres (e.g. techno and house music), and one community has 10 techno DJs, then their genre vector is ${\bf g} = (10,0)$. However, if a community has 5 techno and 5 house DJs, their genre vector is ${\bf g} = (5,5)$. After computing the observed genre vectors for the DJ communities, we measure their pairwise similarity based on the cosine similarity of their genre vectors, as introduced in the main text. These similarity results are shown in Table \ref{tab:sim}.

\begin{table}[htp]
\scriptsize
\begin{tabular}{lllllll}
Paul van Dyk community                                & Armin Van Buuren \& Paul Oakenfold community          & 0.842 &  &  &  &  \\
Sasha \& Tiësto community                             & Carl Cox community                                    & 0.821&  &  &  &  \\
David Guetta community                                & Martin Garrix \& Dimitri Vegas \& Like Mike community & 0.764 &  &  &  &  \\
Sasha \& Tiësto community                             & David Guetta community                                & 0.691 &  &  &  &  \\
Hardwell community                                    & Martin Garrix \& Dimitri Vegas \& Like Mike community & 0.61  &  &  &  &  \\
Sasha \& Tiësto community                             & Armin Van Buuren \& Paul Oakenfold community          & 0.566 &  &  &  &  \\
Carl Cox community                                    & David Guetta community                                & 0.558 &  &  &  &  \\
Carl Cox community                                    & Armin Van Buuren \& Paul Oakenfold community          & 0.489 &  &  &  &  \\
David Guetta community                                & Armin Van Buuren \& Paul Oakenfold community          & 0.444 &  &  &  &  \\
David Guetta community                                & Paul van Dyk community                                & 0.443 &  &  &  &  \\
Hardwell community                                    & David Guetta community                                & 0.414 &  &  &  &  \\
Paul van Dyk community                                & Martin Garrix \& Dimitri Vegas \& Like Mike community & 0.413  &  &  &  &  \\
Paul van Dyk community                                & Sasha \& Tiësto community                             & 0.391 &  &  &  &  \\
Carl Cox community                                    & Paul van Dyk community                                & 0.354 &  &  &  &  \\
Hardwell community                                    & Paul van Dyk community                                & 0.342 &  &  &  &  \\
Martin Garrix \& Dimitri Vegas \& Like Mike community & Armin Van Buuren \& Paul Oakenfold community          & 0.324 &  &  &  &  \\
Hardwell community                                    & Armin Van Buuren \& Paul Oakenfold community          & 0.276 &  &  &  &  \\
Sasha \& Tiësto community                             & Martin Garrix \& Dimitri Vegas \& Like Mike community & 0.224 &  &  &  &  \\
Sasha \& Tiësto community                             & Hardwell community                                    & 0.166 &  &  &  &  \\
Hardwell community                                    & Carl Cox community                                    & 0.166 &  &  &  &  \\
Martin Garrix \& Dimitri Vegas \& Like Mike community & Carl Cox community                                    & 0.138 &  &  &  & 
\end{tabular}
\caption{Cosine similarity of the genre distribution of top DJ communities.}
\label{tab:sim}
\end{table}

\thispagestyle{empty}

\end{document}